\documentstyle[12pt]{article}
\begin{document}
\thispagestyle{empty}
\renewcommand{\thefootnote}{\dagger}

HURD - 9407

March, 1994.
\vskip 2truecm

\begin{center}

{\bf Hilbert - Einstein Action from Induced Gravity coupled with Scalar
Field\\}
\vskip 2truecm

\bigskip
\renewcommand{\thefootnote}{\ddagger}

 {\bf I. L. Shapiro\footnote { e-mail address:
shapiro@fusion.sci.hiroshima-u.ac.jp}}

\bigskip

{\sl Tomsk State Pedagogical Institute, Tomsk, 634041, Russia \\}

{\sl and \\} {\sl Department of Physics, Hiroshima University \\}
{\sl Higashi - Hiroshima, Hiroshima, 724, Japan \\}

\end{center}
\vskip 2.5truecm

\noindent
{\bf Abstract.} Quantum theory of conformal
factor coupled with matter fields is investigated.
The more simple case of the purely classical
scalar matter is considered. It is calculated
 the conformal factor contribution to
the effective potential of scalar field. Then the possibility of the first
order phase transition is explored and the induced values of Newtonian and
cosmological constants are calculated.

\setcounter{page}1
\renewcommand{\thefootnote}{\arabic{footnote}}
\setcounter{footnote}0
\newpage

{\bf Introduction.}
The investigation of quantum effects in extermal gravitational field is one
of the actual ones in a modern gravity and cosmology (see, for example, [1 - 3]
for introduction and references).
There is a hope that the account of quantum effects will give tha way for
solution of some
fundamental problems like the singular structure of classical solutions
and the cosmological constant problem. One of the more important quantum
effects in extermal gravitational field is the appearance of the anomaly trace
of the energy - momentum tensor which allow to calculate, with accuracy to
some conformal invariant functional, the effective action of vacuum [4 - 6].
This effective action originally arise as the nonlocal functional, but can be
written in a local form with the help of some extra dimensionless field
which is named as dilaton in analogy with the string theory, or as conformal
factor. The correlation of this new field with conformal transformation of the
metric have been investigated in [7], where it was proposed to consider the
theory of conformal factor as some kind of infrared gravity model.
In the framework
of the used approximation (all other degrees of freedom except the conformal
factor are frozen) this theory looks very interesting by itself, because it is
superrenormalizable and give possibility to perform an interesting
investigation of the cosmological constant problem. We should like also to
point out some
recent investigation in the field of dilaton gravity [8 - 11].
In particular, in [11] there have been derived the effective potential
and the first order phase transition in the quantum theory of conformal factor.

In fact the introduction of some new kind of fields, like the dilaton field,
is well-defined only after the definition of the interaction between this new
field and the matter fields. Since we hope that the introduction of the
dilaton may solve some physical problems, it is relevant to explore the
possible
ways of it's interaction with matter.
In this paper we start the investigation of the quantum theory of conformal
factor coupled with matter fields. The more simple case of the purely classical
scalar matter is considered. We calculate the conformal factor contribution to
the effective potential of scalar field. Then the possibility of the first
order phase transition is considered  and the induced values of Newtonian and
cosmological constants are calculated.
In contrast to [11] we consider the effective potential for the ordinary
scalar field
and the induced dimensional constants in our case arise as a result of
dimensional transmutation.

{\bf The action of conformal factor and coupling structure.}
The starting point of our investigation is the theory of
asymptotic free and asymptotic conformal invariant massless fields
of spin 0, ${1\over2}$ and $1$ in an external
gravitational field. The asymptotic conformal invariance means that the
value of nonminimal coupling $\xi$ (we suppose that the nonminimal term
$\xi R \phi^2$ is included to the action of scalar field)
have an arbitrary value at low energies
and conformal value $\xi = \frac{1}{6}$ at high energies. We suppose that
at high energies vacuum guantum effects lead to the conformal anomaly
trace of the energy-momentum tensor [12]
$$
T=<T_{\mu}^{\;\;\mu}>=k_1C^2+k_2E+k_3\Box R,   \eqno(1)
$$
where the values of $k_1,_2,_3$ are determined by the
number of fields of different spin. (1) leads to
the equation for the effective action
$$
-{2\over\sqrt{-g}}
g_{\mu\nu}{\delta\Gamma\over{\delta g_{\mu\nu}}}=T.\eqno(2)
$$
which have the following local solution [5]
$$
\Gamma[g_{\mu\nu},\sigma] = S_{c}[g_{\mu\nu}] +
\int d^{4}x \sqrt{-g} \{ {1\over{2}}\sigma\Delta\sigma
+ \sigma[k`_1C^2 + k`_2(E - \frac{2}{3}\Box R)] + k`_3R^2 \}       \eqno(3)
$$
Here $C^2$ is the square of Weyl tensor, $E$ is Gauss - Bonnett invariant, and
conformally covariant operator $\Delta$ is defined as following:
$$
\Delta=\Box^{2}+
2R^{\mu\nu}\nabla_{\mu}\nabla_{\nu}-{2\over{3}}R\Box+{1\over{3}}(\nabla^{\mu}
R)\nabla_{\mu}   \eqno(4)
$$
The values of $k`_{1,2,3}$ differs from $k_{1,2,3}$ because of contribution
of $\Delta$ to conformal trace (1) [5].
The solution (3) contains an arbitrary conformal invariant
funtional $S_{c}$ , which is the integration constant for the
equation (2). This functional is not essential for our purposes and we
shall not take it into account.

Action (3) is just the action of free field $\sigma$ in external metric field,
but the action of Ref.[7] contains the interactions. So the main supposition of
[7] is that the induced gravity and the classical fields appear in a different
conformal points. If we also start with this supposition then it is necessary
to make the conformal transformation of the metric in (3) and then consider
the unified theory. At the same time it is more convinient to make the
conformal transformation of metric and matter fields in the action of the last.
The only source of conformal noninvariance in the action of the fields of
spin $o, \frac{1}{2},1$ is the nonminimal term in scalar sector.
In the framework of asymptotically conformal invariant models the value of
$\xi$ is not equal to $\frac{1}{6}$ at low energies
and hence the interaction of conformal factor with scalar field arise.
Introducing the scale parameter $\alpha$ we obtain the following action
of conformal factor coupled with scalar field.
$$
S = \Gamma[g_{\mu\nu},\sigma] +
\int d^{4}x \sqrt{-g} \{ ({1\over{2}}g^{\mu\nu}\partial_{\mu}\phi
\partial_{\nu}\phi + {1\over{2}}\xi R \phi^2 +
$$
$$
+{1\over{2}}(1 - 6\xi)\phi^2 [\alpha^2
g^{\mu\nu}\partial_{\mu}\sigma \partial_{\nu}\sigma + \alpha (\Box \sigma)]
- \frac{\lambda}{4!}\phi^4.        \eqno(5)
$$
So the interaction between scalar field and conformal factor arise as a result
of conformal thansformation of the metric
and matter fields
$$g_{\mu\nu} \rightarrow g`_{\mu\nu} = g_{\mu\nu} exp(2\alpha\sigma(x))\;\;\;
\Phi\rightarrow \Phi` = \Phi exp(d_{\Phi}\alpha\sigma(x))$$
where $d_{\Phi}$ is conformal weight of the field
$\Phi$. The only kind of fields which take part in such an interaction is
scalar
 one, where the interaction with conformal factor appears as a result of
nonconformal coupling at low energies. In the next section we shall discuss
the renormalization of the theory (5) and also derive the effective potential.

{\bf Renormalization and effective potential.} Since the metric and scalar
field are frozen, (5) is just the action of free field $\sigma$ in external
metric and scalar fields. Therefore the effective action of the theory (5) is
given by expression
$$
W = \frac{i}{2}Tr\ln\hat{H}  \eqno(6)
$$
 where $\hat{H}$ is Hermitian bilinear form of the action (5).
$$
\hat{H} = \Box^2 + V^{\mu\nu}\nabla_{\mu}\nabla_{\nu} + N^{\mu}\nabla_{\mu}+U
\eqno(7)
$$
where
$$
V^{\mu\nu} = 2R^{\mu\nu} - \frac{2}{3}Rg^{\mu\nu} +
(6\xi - 1)\phi^2 \alpha^2 g^{\mu\nu}
,\;\;\;U = 0.   \eqno(8)
$$
Note that the value of $N^{\mu}$ is not essential for our purposes.
The divergent part of (6) can be easily obtained by standard methods [13],
and we get (omitting surface terms):
$$
 W_{div} = -\frac{1}{\varepsilon}
\int d^{4}x \sqrt{-g} \{ -\frac{2}{15}(R^{\mu\nu}R_{\mu\nu} - \frac{1}{3}R^2)
+ {1\over{2}}\alpha^4(1 - 6\xi)^2 \phi^4 \}
      \eqno(9)
$$
{}From (9) follows that parameter $\lambda$ is the only one which have to be
renormalized. Here we do not consider the renormalization of vacuum and
surface parts. The $\beta$ - function corresponding to $\lambda$ have the form
$$
\beta_\lambda = 12 \alpha^4(1 - 6\xi)^2. \eqno(10)
$$
It is interesting that the effective coupling  $\lambda(t)$ have asymptotically
free behaviour in IR limit, just as in ordinary $\lambda\phi^4$ theory.
One can consider this fact as another evidence in favour of IR character of
the theory (5). The effective potential can be calculated (with the accuracy to
the first order in curvature) in the framework of
standard methods [14] (see also [3]). If we use the normalization conditions
of [3], we obtain
$$
V_{eff} = - {1\over{2}}\xi R \phi^2 + \frac{\lambda}{4!}\phi^4
+ {1\over{4}}\alpha^4(1 - 6\xi)^2 \phi^4 \} \left[ \ln\frac{\phi^2}{\mu^2}
- \frac{25}{6} \right],      \eqno(11)
$$
where $\mu$ is dimensional parameter of renormalization.

{\bf Phase transition induced by curvature.}
We shall restrict ourselves by the consideration of the only first - order
phase transition. Then the equations for the critical values of curvature
$R_c$ and order parameter $\phi_c$ have the form:
$$
V(R_c,\phi_c) = 0, \;\;\;\;\; V`(R_c,\phi_c) = 0,
\;\;\;\;\;V``(R_c,\phi_c) > 0.
\eqno(12)
$$
Here primes stand for derivatives of $V$ with respect to $\phi$.
The first two equations in (12) give the following expressions for critical
values of curvature and $\phi$.
$$
\phi_c^2 = p|R_c|, \;\;\;\;\;\; p = - \frac{2\varepsilon\xi}{\alpha^4(1 -
6\xi)^2}
\eqno(13)
$$
$$
|R_c| = p^{-1} \mu^2 exp \left[ \frac{19}{6} - \frac{\lambda}
{6\alpha^4(1 - 6\xi)^2} \right]   \eqno(14)
$$
where $\varepsilon$ is sign of $R_c$, which is necessary negative in this
theory. The third condition (12) reads as
$\alpha^4(1 - 6\xi)^2 > 0$. and is fulfilled. Substituting (13), (14) into
effective potential (11) we obtain the estimate for the value of $V_{eff}$ in
the critical point. The corresponding action have the form of Hilbert -
Einstein action
$$
S_{ind} = - \frac{1}{16\pi G_{ind}} \int d^{4}x \sqrt{-g} (R - 2\Lambda_{ind})
\eqno(15)
$$
where induced values of Newtonian and cosmological constants are defined from
(11) - (14).
$$
\frac{1}{16\pi G_{ind}} = - \xi\mu^2 exp \left[ \frac{19}{6} - \frac{\lambda}
{6\alpha^4(1 - 6\xi)^2} \right]   \eqno(16)
$$
$$
\Lambda_{ind} = - \frac{1}{16\pi G_{ind}} \frac{\alpha^4(1 - 6\xi)^2}{16\xi^2}
	                                   \eqno(17)
$$
The expressions (16),(17) contains the arbitrariness related with the value
of $\mu$. One can identify $\mu$ with the point on mass scale, which
corresponds to the phase transition [15]. Then we can get the unambigous
values of induced quantities.

{\bf Discussion.} Let now give brief analysis of (16),(17). It is easy to see,
that one can get a tiny value of the induced cosmological constant if taken
the value of $\alpha$ to be small enough. Just the same result appears when
$\xi$ is close to conformal value $\frac{1}{6}$. In fact we suppose that the
energy scale of strong matter effects in external gravitational field is some
$\mu_1$ and that the quantum effects of conformal factor is relevant at another
scale $\mu_2 < \mu_1$. Since the close curvature and energy scales correspond
to a little values of scale factor $\alpha$, we obtain, that if the difference
between $\mu_2, \mu_1$ is little enough, the induced value of cosmological
constant is small. Note that the value $\xi \approx \frac{1}{6}$ also
corresponds to the case of close energy scales $\mu_1$ and $\mu_2$. So if the
quantum effects of conformal factor are taken into account we get an effective
and natural mechanism of fine - tuning for the cosmological constant.

The above analysis is rather simplified due to the absence of quantum effects
of matter fields. In fact the only nontrivial contributions to $V_{eff}$ come
from the quantized scalar field. All other fields decouple from conformal
factor and therefore give an additive contributions to $V_{eff}$. However
the appearance of quantized scalar field leads to the nontrivial  mixed sector
in the bilinear form of the action and the renormalizability can be broken.
In our opinion this question have to be investigated separately. The next
intersting problem is incorporation of quantum metric. Then it is
necessary to consider the general dilaton gravity [8,9,11] which is quite a
 direct
generalization of higher derivative gravity theory [16 - 19, 13].  In this
case the interaction between gravity (including conformal factor) and matter
fields takes place even without the conformal shift of the induced action.
Generally speaking such an investigation is possible, at least, on one - loop
level, because the technique of [20, 3] looks applicable to this case (see,
also, [21]), but it
is related with a very involved calculations. So one can consider the
problem explored  above, as some toy model for the general dilaton gravity
coupled with matter fields and the main result as some kind of indication
to the possible solution of the cosmological constant problem.

 \end{document}